# Ontologies for Network Security and Future Challenges

**Danny Velasco Silva[1] and Glen Rodríguez Rafael[2]**
[1]**University National of Chimborazo, Ecuador**
[1, 2]**University National Mayor of San Marcos, Perú**
dvelasco@unach.edu.ec
glen.rodriguez@gmail.com

**Abstract:** Efforts have been recently made to construct ontologies for network security. The proposed ontologies are related to specific aspects of network security. Therefore, it is necessary to identify the specific aspects covered by existing ontologies for network security. A review and analysis of the principal issues, challenges, and the extent of progress related to distinct ontologies was performed. Each example was classified according to the typology of the ontologies for network security. Some aspects include identifying threats, intrusion detection systems (IDS), alerts, attacks, countermeasures, security policies, and network management tools. The research performed here proposes the use of three stages: 1. Inputs; 2. Processing; and 3. Outputs. The analysis resulted in the introduction of new challenges and aspects that may be used as the basis for future research. One major issue that was discovered identifies the need to develop new ontologies that relate to distinct aspects of network security, thereby facilitating management tasks.

## 1. Introduction

Computer network security has steadily become a very extensive field of research. Networking and networks have opened new horizons that allow us to explore beyond the boundaries of current institutions. This situation has led to the emergence of new threats to computerized systems. The domain of network security encompasses a set of methods, techniques and tools responsible for protecting the resources and guaranteeing the availability, confidentiality, integrity and the traceability of information. This knowledge gathered from the information security domain can be formally described by ontologies, which enables the modelling of network security.

According to Gruber (1995), an ontology is an explicit specification of a conceptualization. A conceptualization is an abstract and simplified view of the reality portion that interests us: objects, concepts and other units that exist in some area of interest and the relationships among them. Together with the benefits contributing to the use of complete ontologies that preserve the handling of all concepts in a specific area, this notion guides us to the knowledge of the existing proposals within ontological engineering, focusing on the area of network security and identifying the problems affecting the management of large-scale networks. From the conducted studies regarding ontologies in network security, those that cover certain aspects but are not aimed at solving security requirements, defence strategies, assets, data integration in using tools, and security protocols, which are necessary for a better network management, can be identified. There have been efforts by authors focused on analyzing specific aspects of network security. This has identified the need to formulate a new model, a base ontology that encompasses all types of network security ontologies. It must be formulated by analyzing the potentialities of studies carried out in the field, reusing well-defined reference works, overcovering the shortcomings of these works, and incorporating aspects not considered for the purpose of having centralized information. It must also outline an improved management structure to make correct and timely decisions that maintain network security.

The process in this research was performed using the analysis of the main identified ontologies proposed by Levy and Ellis (2009) to propose a comprehensive ontology.

The paper is organized as follows:

- Section II gives the methodology used for the development of this research,
- Section III shows the literature review,
- Section IV shows the results and discussion,
- Section V displays the proposal, and
- Section VI provides the conclusions and future research directions.

## 2. Methodology

The three suggested steps are as follows:

1. Inputs: The literature was obtained from different databases, and research was conducted using the following phrase: Ontology Network Security.

2. Processing: From the relevant studies acquired in the literature for the selected research, the developed activities were as follows: summarize, differentiate, interpret and contrast the documents that help address the posed research question: what aspects have been covered by the ontologies in network security?

3. Outputs: The selected articles were classified according to their applicability; the results of this phase are shown in the following section.

## 3. Literature review

Several research papers have been published regarding the construction of ontologies in the field network security. Some efforts have focused on the development of security ontologies to prepare for threats, IDS, alerts, attacks, vulnerabilities, countermeasures, security policies, network management.

Undercoffer, Joshi, and Pinkston (2003), stated the benefit of transitioning from taxonomies to ontologies and proposed an ontology to simulate computer attacks for sharing knowledge of intrusion detection systems. The authors used DAML+OIL and DAMLJessKB to implement the ontology and present use case scenarios to illustrate the benefits of employing the ontology.

Undercoffer et al. (2004) presented an ontology for computer network attacks that they have used for distributed IDS. The authors analyzed four thousand classes of computer network attacks, and examined their relationships to each other and their attributes.

Razzaq et al. (2009) presented an intelligent system with an ontological basis that analyzes the input semantically and is able to detect attacks with irrelevant rates of false positives.

Gyrard, Bonnet, and Boudaoud (2013) proposed a security ontology that defines the key security concepts such as attacks, countermeasures, security properties, and their relation to each other.

Gao et al. (2013), provided an ontology-based attack model, and then utilized it to assess the information systems' security from the attack perspective. They categorized attacks into a taxonomy convenient for security assessment. The proposed taxonomy consisted of five dimensions, which include attack impact, attack vector, attack target, vulnerability, and defense.

Simmons, Shiva, and Simmons (2014) proposed an ontology-based problem-solving system used to identify and defend against cyber attacks.

Razzaq et al. (2014) demonstrated how an ontological engineering methodology can be administered systematically to design and evaluate security systems. It shows a detailed ontological model that satisfies the broad utilization of web applications, communication protocols, and attacks.

Si et al. (2014) constructed a fusion model comprised of class keys such as network environment, network vulnerability, network attack, network security incident, and sensors. In addition, they formulate three fusion rules that contain the aggrupation and verification of alerts and the reconstruction of attack sessions.

Karande and Gupta (2015) proposed an IDS ontological model that detects protocol-specific attacks and identifies malicious scripts. This model identifies types of attacks and vulnerabilities.

Wang and Guo (2009) proposed an ontological approach to capture and utilize the fundamental concepts of information security and their function to recover vulnerability data and make inferences about the cause and impact said data.

Elahi, Yu, and Zannone (2009) proposed a modeling ontology focused on vulnerability. It aimed to integrate empirical knowledge of vulnerabilities into the system development process.

Wang and Guo (2009) examined using logic to apply semantic technology to information security with a focus on software vulnerability management.

Bhandari and Gujral (2014) presented an ontological approach to perceiving the current security state of the network. Vulnerabilities and attacks are the main classes of taxonomy in the ontology.

Do Amaral et al. (2006) presented a ontological structure for information security and considered a paradigm through which it can be used to extract knowledge from natural language texts such as information security standards, security policies, and security control descriptions.

Xu et al. (2008) discussed the potential of applying an integration of ontology-based and policy-based approaches to automate pervasive network security management, and then proposes a model in order to validate the feasibility of this integrated approach.

Cuppens-Boulahia et al. (2009) proposed an ontology based approach to present instances of security policies.

Ramanauskaitė et al (2013) analyzed existing security ontologies by comparing their general properties, using OntoMetric factors with the ability to satisfy different security standards.

Ye, Bai, and Zhang (2008) presented the design and implementation of a representation based on the knowledge of ontology for a distributed multi-agent peer-to-peer IDS.

Colace, De Santo, and Ferrandino (2012) introduced a Network Intrusion Prevention System based on Ontological and Slow Intelligence approach.

Khairkar, Kshirsagar, and Kumar (2013) addressed the problems of existing IDS software. Regarding the index of false positives and false negatives, as well as data overloads, a proposition was proposed that included using the concepts of the semantic web and ontologies with the purpose of defining an approach for analyzing security logs in order to identify possible security issues.

Sartakov (2015) presented an ontological representation of a network to create a specification-based IDS.

Kyriakopoulos, Parish, and Whitley (2015) introduced a web-based tool for network management to control a large number of network monitoring data sources, using artificial intelligence through an intuitive user interface.

Lannacone et al. (2015) described an ontology developed from a database of cyber security knowledge graphs. It is intended to provide an organized framework that incorporates information from a variety of structured and unstructured data sources.

Xu, Xiao, and Wu (2009) proposed the use of a security ontology that consolidates knowledge and information in order to perform a contextual alert analysis and analyze in detail the problems of constructing the security ontology using OWL + SWRL + OWL-S.

Li and Tian (2010) focus on how to develop an intrusion alert correlation system. The system is based on a hierarchical model of alert correlation knowledge and the XSWRL ontology technique.

The majority of the articles reviewed base their proposals on a specific methodology. In addition, they use tools for the construction of the ontology. The most frequently used tool is Protégé, a free, open-source ontology publisher and framework for building intelligent systems.

Ye, Bai and Zhang (2008), Xu et al. (2008), Xu, Xiao and Wu (2009), Colace, De Santo and Ferrandino (2012), Si et al. (2014), and Razzaq et al. (2014) evaluated their proposals through test scenarios.

Xu et al. (2008), Simmons, Shiva and Simmons (2014), Razzaq et al. (2014), and Karande and Gupta (2015) validated their proposals using a qualitative study by professional security experts.

## 4. Results and discussion

To present the achieved results, previous studies were grouped according to their individual contributions. The purpose was to understand the trends in the field, as indicated in Table 1 below.

**Table 1:** Network security ontologies

| Type of Publication | Source | Author | Aspects | | | | | | | |
|---|---|---|---|---|---|---|---|---|---|---|
| | | | Threats | IDS | Alerts | Attacks | Vulnerabilities | Countermeasures | Security policies | Network management |
| Journal | Science Direct | Li and Tian (2010) | ○ | ○ | ● | ● | ● | ○ | ○ | ○ |
| | | Gao et al (2013) | ● | ○ | ○ | ● | ● | ● | ○ | ○ |
| | | Razzaq et al (2014) | ○ | ○ | ○ | ● | ● | ○ | ● | ○ |
| | | Ramanauskaitė et al (2013) | ● | ○ | ○ | ○ | ● | ● | ○ | ○ |
| | ACM Digital Library | Cuppens-Boulahia et al. (2009) | ● | ○ | ● | ● | ○ | ○ | ● | ○ |
| | Taylor & Francis | Wang, Guo and Camargo (2010) | ○ | ○ | ○ | ● | ● | ● | ○ | ○ |
| Conference | ACM Digital Library | Undercoffer et al (2004) | ○ | ○ | ○ | ● | ○ | ○ | ○ | ○ |
| | | Wang and Guo (2009) | ○ | ○ | ○ | ○ | ● | ○ | ○ | ○ |
| | | Gyrard, Bonnet and Boudaoud (2013) | ○ | ○ | ○ | ● | ○ | ● | ○ | ○ |
| | | Lannacone et al (2015) | ○ | ○ | ○ | ● | ● | ○ | ○ | ○ |
| | IEEE Xplorer | Do Amaral et al (2006) | ○ | ○ | ○ | ○ | ○ | ○ | ● | ○ |
| | | Xiao and Xu (2006) | ○ | ○ | ○ | ○ | ○ | ○ | ● | ○ |
| | | Xu et al (2008) | ○ | ○ | ○ | ○ | ○ | ○ | ● | ○ |
| | | Ye, Bai and Zhang (2008) | ○ | ● | ○ | ● | ○ | ○ | ○ | ○ |
| | | Razzaq et al (2009) | ○ | ○ | ○ | ● | ○ | ● | ○ | ○ |
| | | Wang and Guo (2009) | ○ | ○ | ○ | ○ | ● | ○ | ○ | ○ |
| | | Xu, Xiao and Wu (2009) | ○ | ○ | ● | ○ | ● | ● | ○ | ○ |
| | | Colace, De Santo and Ferrandino (2012) | ○ | ● | ○ | ○ | ○ | ○ | ○ | ○ |
| | | Frye, Cheng and Heflin (2012) | ○ | ● | ○ | ● | ● | ○ | ○ | ○ |
| | | Khairkar, Kshirsagar, and Kumar (2013) | ● | ● | ○ | ● | ○ | ○ | ○ | ○ |
| | | Bhandari, and Gujral (2014) | ○ | ○ | ○ | ● | ● | ○ | ○ | ○ |
| | | Lundquist, Zhang and Ouksel (2014) | ● | ○ | ○ | ○ | ○ | ○ | ○ | ○ |
| | | Si et al (2014) | ○ | ○ | ● | ● | ● | ○ | ○ | ○ |
| | | Simmons, Shiva and Simmons (2014) | ○ | ○ | ○ | ● | ○ | ○ | ○ | ○ |
| | | Kyriakopoulos, Parish and Whitley (2015) | ○ | ○ | ○ | ○ | ○ | ○ | ○ | ● |
| | | Karande and Gupta (2015) | ○ | ● | ○ | ● | ● | ○ | ○ | ○ |
| | Springer | Undercoffer Joshi and Pinkston (2003) | ○ | ○ | ○ | ● | ○ | ○ | ○ | ○ |
| | | Sandilands, and Van Ekert (2004) | ● | ○ | ○ | ● | ● | ○ | ○ | ○ |
| | | Elahi Yu, and Zannone (2009) | ○ | ○ | ○ | ● | ● | ○ | ○ | ○ |
| | | Sartakov (2015) | ○ | ● | ○ | ○ | ○ | ○ | ○ | ○ |
| | | | 6 | 7 | 4 | 19 | 15 | 6 | 5 | 1 |

A large component of the proposed work includes ontologies that focus on specific aspects regardless of the entire network security domain. The relative percentages of the studied aspects are shown in Figure 1 below.

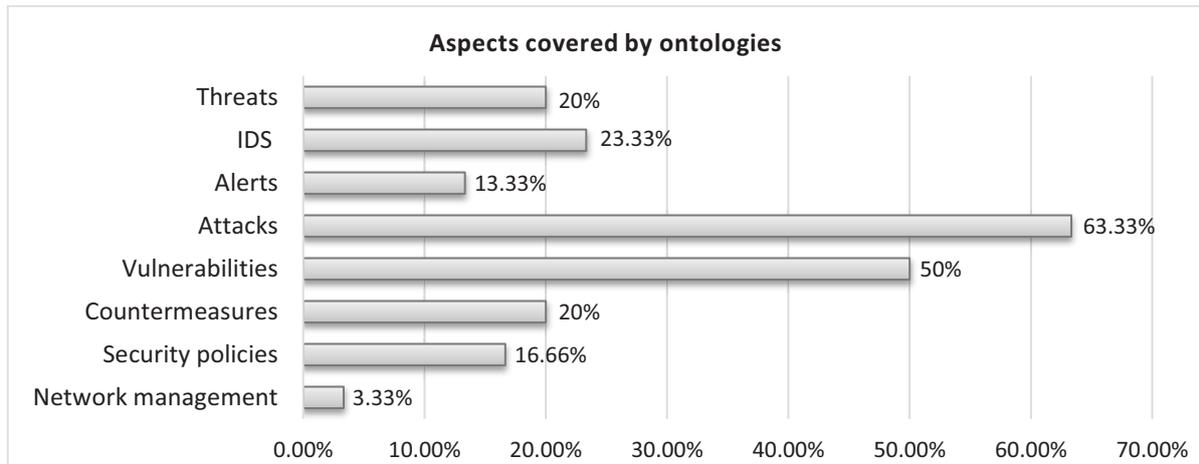

**Figure 1:** Aspects covered by ontologies

63.33% of the ontologies make reference to attacks and their taxonomical structure. Their focus is mainly on the network layer missing attacks at the application layer.

80% of the papers reviewed do not present the results obtained from test scenarios, and therefore it is unachievable to evaluate the ontology and determine if it adapts to the requirements or to measure its effectiveness. Only 13.33% of the papers validate their proposals, trying to identify the correct use of the language, the accuracy of the taxonomic structure, the validity of the vocabulary, and the adequacy of the requirements for the purpose of documenting the process of development to verify if the proposal complies with the terms specified.

One of the challenges that constitutes a potentially interesting area arises when data is collected from different safety equipment (IDS, Intrusion prevention system, firewall, antivirus system, system security audit, honeynet, etc.). The safety equipment is distributed in different domains in the network, which is required to develop an ontology that can integrate real-time data from this safety equipment and allows the captured data to be properly administered.

## 5. Proposal

In Figure 2, an integral ontology is proposed that covers aspects that were not considered in prior works and provides an improved management to make correct and timely decisions that maintain network security.

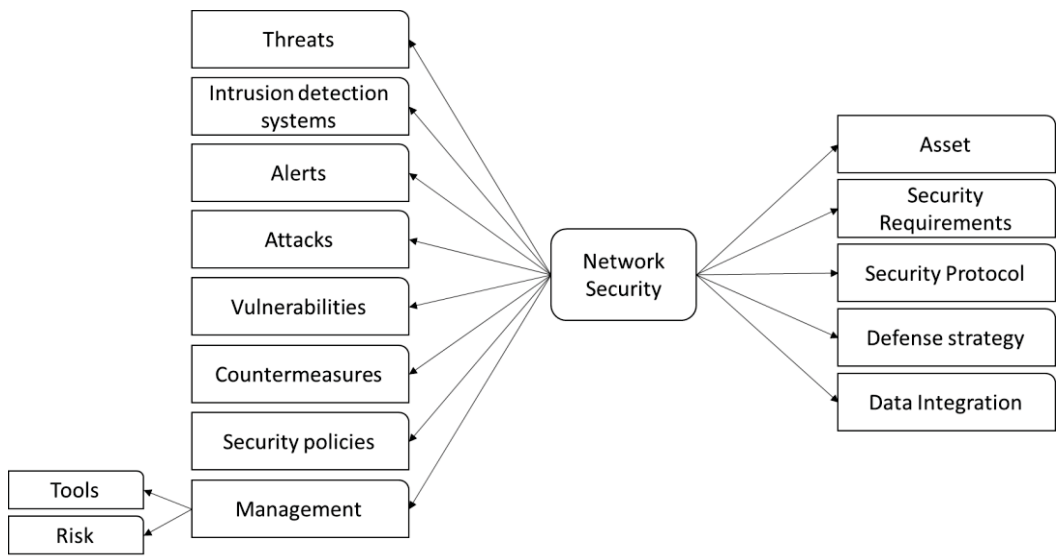

**Figure 2:** Comprehensive ontology in network security

A comprehensive ontology that covers the network security domain is suggested. It is advantageous because it has a formal specification, facilitates knowledge sharing, and reuses the considered aspects in the particular domain of knowledge. In studies concerning the ontologies, there are aspects that have not been considered and are part of this proposal, such as defining security strategies to mitigate the threats that may be exposed on a computer network; considering security requirements that are generally proposed from engineering safety, specific concepts and measures; protecting the assets, which can be information, hardware, or software, that have value to the organization and can be targeted by attackers; having security protocol that contribute to having a more robust infrastructure; and integrating data that support the information integration theory, which was proposed to describe and model how information from different sources is integrated to make an overall judgement, leading to a single format presentation of information provided by the network monitoring.

## 6. Conclusions and future work

A large part of the proposed ontologies are focused on covering specific aspects. Consequently, the security community needs an ontology that covers the entire network security domain that is flexible, adaptable, and facilitates the reuse, communication and exchange of knowledge by providing network scalability. This is accomplished using mapping techniques to integrate all of the specific aspects, focusing on the general requirements for network security and availability, authentication, integrity and confidentiality.

63.33% of the ontologies reviewed focused on attacks and 50% focused on vulnerabilities, which reflects the thinking that attacks and vulnerabilities are the most preponderant aspects. However, they neglect other aspects that are included in the domain of network security.

Only 20% evaluated their proposals in test scenarios, which allow us to discern the errors in a system before its implementation, favoring construction through feedback. 13.33% validated its proposals through security experts. It is necessary for these mechanisms of evaluation and validation to look for more effective methods of analysis of ontological systems.